\documentclass[fdp,a4paper,fleqn%
]{w-art}

\newcommand{\dd}{\mathrm{d}}
\newcommand{\e}{\mathrm{e}}
\newcommand{\w}{\wedge}
\newcommand{\bbm}{\left(\begin{matrix}}
\newcommand{\ebm}{\end{matrix}\right)}
\newcommand{\beq}{\begin{equation}}
\newcommand{\eeq}{\end{equation}}

\makeatother

\usepackage{times,cite,w-thm}

\usepackage[]{graphicx}
\begin{document}

\pagespan{1}{}




\begin{flushright}
ITP-UH-04/14
\end{flushright}

\title[U-dual branes]{U-dual branes and mixed symmetry tensor fields}


\author[A. Chatzistavrakidis]{Athanasios Chatzistavrakidis\inst{1,}%
 \footnote{ \textsf{thanasis@itp.uni-hannover.de},
            }}
\address[\inst{1}Institut f{\"u}r Theoretische Physik]{Appelstra{\ss}e 2, 30167 Hannover, Germany}
\author[F. F. Gautason]{Fridrik Freyr Gautason\inst{1,2,}\footnote{\textsf{fridrik.gautason@itp.uni-hannover.de}}}
\address[\inst{2}Center for Quantum Engineering and Spacetime Research]{Appelstra{\ss}e 2, 30167 Hannover, Germany}
\begin{abstract}
We review and explain the relation between U-dual branes in string theory and mixed symmetry tensors of various degrees. In certain cases
these mixed symmetry tensors can be related to diverse types of fluxes that play an important role in compactifications of string theory.
\end{abstract}
\maketitle                   





String dualities have proved a useful tool in understanding several non-perturbative aspects of string theory. 
One of its consequences is that the interplay between T- and S-dualities and D-branes leads to many new states, 
which were classified and studied from different points of view in Refs. \cite{exotic1,exotic2,defect1,defect2,axel}.
A fraction of these branes, dubbed exotic, was associated to non-geometric backgrounds in Ref. \cite{exotic1} and this fact was 
studied further in Ref. \cite{Chatzistavrakidis:2013jqa}, where this contribution is partly based. Other closely 
related recent work may be found in Refs. 
\cite{Hassler,betz,Kimura:2014upa}. 
Here, due to lack of space, our purpose is to emphasize the nature of the background fields that couple to exotic 
branes and to provide more precise explanations than Ref. \cite{Chatzistavrakidis:2013jqa}
about the role of mixed symmetry tensor fields in this context. 

 Let us begin with the following five-branes that exist in type II superstring theory:
\beq\label{branechain}
\text{D5} \quad\overset{S}{\longleftrightarrow}\quad  \text{NS5} \quad \overset{T_y}{\longleftrightarrow} \quad\text{KKM} 
\quad \overset{T_z}{\longleftrightarrow} \quad 5^2_2 \quad\overset{S}{\longleftrightarrow}\quad 5^2_3~,
\eeq
where the duality relations among them are also manifest.
This diagram depicts five different string backgrounds that, assuming the U-duality symmetry of the type II superstring  
as suggested by Hull and Townsend \cite{hulltownsend}, are physically equivalent. Reading from left to right we first encounter the standard type IIB D5-brane which couples magnetically to the RR gauge potential $C_2$, where the subscript 
denotes the form degree of the field. The NS5-brane is related to the D5-brane via S-duality 
and it couples magnetically to the NSNS gauge potential $B_2$. By performing a T-duality of the NS5-brane background
along an isometry direction, labelled $y$ for concreteness, the type IIA Kaluza-Klein Monopole (KKM) is obtained.
The KKM is an example of a brane whose existence can only be inferred from the dimensionally reduced theory.
In the dimensionally reduced theory one finds that the KKM acts as a magnetic source for the KK vector, which we denote by $A_1^y$. 
The latter is an elementary mixed symmetry tensor, being an 1-form and carrying a vector index too. 
A further T-duality along a second isometry direction (hence labelled $z$) leads to the so-called $5^2_2$-brane. In this 
notation for branes the subscript denotes the power of $1/g_s$ in the brane's tension and the superscript denotes the number of special NUT-like directions associated with the brane. In this way we can also denote the D5 as $5_1$, 
the NS5 as $5_2$ and the KKM as $5_2^1$. By performing the double T-duality explicitly one finds that the $5^2_2$-brane is a magnetic source of an eight-dimensional scalar $\beta^{yz}$ with two vector indices, which we will revisit in a moment.
The $5_2^2$-brane can be related to the $5_3^2$ one via the type IIB S-duality; the latter acts as a magnetic source 
for a dual scalar, say $\gamma^{yz}$, with two vector indices in eight dimensions. 

A diagram similar to (\ref{branechain}) is often encountered in the flux compactification literature where the entries correspond to 
certain flux compactifications,
\beq\label{fluxchain}
H_3\quad  \overset{T_y}{\longleftrightarrow}\quad f_2^1 \quad\overset{T_z}{\longleftrightarrow} \quad Q_1^2~,
\eeq
where we denote the tensor fields with a subscript indicating their form degree and a superscript indicating their vector degree. 
Note that this notation should not be confused with the similar one used for the branes.
In particular, $H_3$ is the Kalb-Ramond field strength and it is a pure form of degree 3. It corresponds to 
the familiar geometry of a flat torus threaded by $H_3$ flux.
The rest of the objects always 
involve some vector indices. 
 For the second entry no \emph{standard} gauge flux is present but the torus is not flat, having geometric torsion 
 (or geometric flux) denoted by $f_2^1$. This is in full agreement with the torsion tensor, which is a vector-valued 2-form.
 A second T-duality relates this background to the non-geometric $Q$-background.  The terminology \emph{non-geometry}
 refers to the fact that the metric and the gauge potential $B_2$ are not globally well-defined in this case.
 More precisely these background fields cannot be glued along different patches of the torus without using 
 duality transformations. 
 By now we understand that such a situation finds an elegant interpretation in the framework of generalized complex geometry \cite{Hitchin:2004ut,Gualtieri}. There one defines the generalized metric as a combination of the metric $G$ and the $B_2$ field
\beq\label{generalizedmetric}
\cal H = \left(\begin{array}{cc} G-B_2G^{-1}B_2  & B_2G^{-1} \\ 
-G^{-1}B_2 & G^{-1} \end{array}\right)~,
\eeq
and finds that it is well-defined. This implies that a certain field redefinition can be made that yields a well-defined metric $g$, 
different than $G$,
and an accompanying 2-vector $\beta^2$. The field $\beta^2$ has a field strength $Q_1^2$, a 2-vector-valued 1-form, which explains the entry in diagram \eqref{fluxchain}. By extending the chain \eqref{fluxchain} in both directions with S-duality,
\beq\label{extfluxchain}
F_3 \quad \overset{S}{\longleftrightarrow} \quad H_3 \quad \overset{T_y}{\longleftrightarrow}\quad  f_2^1\quad \overset{T_z}{\longleftrightarrow}\quad Q_1^2\quad\overset{S}{\longleftrightarrow} \quad P_1^2~,
\eeq
one readily observes the analogy with the diagram \eqref{branechain}. Here $F_3$ is the 3-form field strength of the 
RR gauge potential $C_2$, while $P^2_1$ is another 2-vector-valued 1-form, corresponding to the dual 2-vector that we denoted as 
$\gamma^{yz}$. We conclude that the branes in \eqref{branechain} are the magnetic sources for the fluxes in \eqref{extfluxchain}. In Ref. \cite{Chatzistavrakidis:2013jqa} it was explicitly shown
that the $5_2^2$-brane is coupled to the magnetic dual of $\beta^2$, which we denote by $\beta_8^2$. 
At the linearized level this magnetic dual is defined by
\beq\label{magneticdual}
\dd \beta_8^2 =  \e^{-2\phi} \star \dd \beta^2~,
\eeq
where $\phi$ is the dilaton and $\star$ is the standard Hodge duality operator. Notice that the operation of dualizing 
the field $\beta^2$ does not see the vector indices. This is in line with the way one dualizes mixed symmetry forms, 
which we will discuss shortly. First we note that although the magnetic dual of $\beta^2$ is naively an 8-form, the $5^2_2$-brane is 
still a five-brane, as the name suggests. It was noted in Refs. \cite{defect1,defect2,Chatzistavrakidis:2013jqa} that in order to write down the Wess-Zumino action for such a brane, the vector indices of $\beta_8^2$ and two of its form indices should be 
explicit and equal, namely $\beta^{yz}_{i_1\dots i_6yz}$. Then the leading term in the Wess-Zumino coupling is proportional to 
\beq\label{wz}
\int_{{\mathcal M}_6} \iota_y\iota_z \beta_8^{yz}~,
\eeq
where $\iota_y$ ($\iota_z$) is the contraction in the isometry direction $y$ ($z$) and ${\mathcal M}_6$ is the 
brane world volume. Since this is a point of primary importance 
for our discussion, let us provide a more thorough argument for it. 
Since the field strength $Q_1^2$ is dual to a pure $H_3$ flux from a lower-dimensional viewpoint, 
this means that it vanishes whenever its single form index is the same with either one of its two vector indices.
This directly implies that the Hodge dual $Q_9^2 = \e^{-2\phi}\star Q_1^2$ necessarily has two of its nine form indices
equal to its two vector indices. Furthermore, since $y$ and $z$ are isometry directions the form index of $Q_1^2$ 
cannot be either $y$ or $z$. This guarantees that two form indices of $Q_9^2$ must always be $y$ and $z$.
The same line of arguments holds for $\beta_8^2$ too because the exterior derivative cannot be in one of the isometry directions.
It follows that $\beta_8^2$ only carries the degrees of freedom of a 6-form when the vector indices are made explicit in the isometry direction and the integration in Eq. (\ref{wz}) indeed makes sense.

Let us discuss in more detail the properties of mixed symmetry forms. A mixed symmetry form \cite{mixed1,mixed2,mixed3} is a tensor with two or more sets of antisymmetrized (or form) indices. In that sense, a general such form can be written as 
\beq 
\omega_{p,q}=\frac {1}{p!q!}\omega_{i_1\dots i_pj_1\dots j_q}\dd x^{i_1}\wedge\dots\wedge \dd x^{i_p}\otimes \dd x^{j_1}\wedge\dots\wedge\dd x^{j_q}~,
\eeq 
in the case of two sets of antisymmetrized indices
and similarly for cases with more such sets. Its degree as a tensor is obviously $p+q$ and hereby we denote it as $[p,q]$. In 
standard tensor notation its degree is simply $(0,p+q)$, where the first entry counts vector indices and the second entry form indices.
In this contribution we focus on mixed symmetry forms with two sets of indices, although more involved cases 
are not irrelevant. 
A $[p,q]$-form as above can be represented by a Young tableaux where in our case the tableaux only contains two columns and the columns have length $p$ and $q$ respectively. Sometimes a trace condition is assumed, whence the contraction of any index from the first set with any index from the second set gives zero \cite{mixed3}. This is essentially a similar condition to the one for $Q^2_1$ and $Q^2_9$ that we mentioned and 
argued for above, but for the moment we do not require this condition. 
In Ref. \cite{mixed3} the exterior derivative comes in two types, one for each set of indices. A similar doubling occurs for 
the Hodge operator. Although there is nothing wrong with this, here we use mixed symmetry tensors in a slightly different sense. In particular, we examine standard tensor fields with one set of antisymmetrized form indices and one set of 
antisymmetrized vector indices, i.e.
\beq 
\omega_p^q=\frac 1{p!q!}\omega_{i_1\dots i_p}^{j_1\dots j_q}\dd x^{i_1}\wedge\dots\wedge \dd x^{i_p}\otimes \partial_{j_1}\wedge \dots 
\wedge \partial_{j_q}~.
\eeq
From now on we will use the standard notation $(q,p)$ to denote a mixed symmetry tensor of this type, where $q$ of the $q+p$ indices are vector indices. We have already seen several objects that fall under the above definition of mixed symmetry tensor fields, such as
$
f_2^1, Q_1^2,\beta_8^2
$ and others.
Comparing to the previous definition, the difference now is that there is only one exterior derivative and only one 
Hodge operator, both acting on the lower (form) indices.

The relation between non-standard branes and mixed symmetry forms was first suggested in Refs. \cite{defect1,defect2}. Using the representation theory of the U-duality group it is concluded that more than one magnetic dual must exist for all gauge potentials of the type II 
superstring. This is due to the fact that this group-theoretical analysis reveals more brane states than could be accounted for
by the standard NSNS and RR sector together with the magnetic duals of the gauge potentials. 
The existence of those additional duals allows the $B_2$ field to couple not only 
to the NS5-brane via the standard magnetic dual $B_6$, but also to the $5^2_2$-brane via an exotic magnetic dual. 
The latter is exactly the mixed symmetry form $\beta_8^2$ that we discussed before. We can also think of this in a different way,
 using expressions from generalized geometry to exchange the $B_2$ field for a 
2-vector $\beta^2$ which has a magnetic dual $\beta_8^2$, cf. Eq. (\ref{magneticdual}). 
Similar considerations apply to the KKM. The frame field or vielbein $e_1^1$ can be thought of as a vector-valued 1-form or
 as a mixed symmetry tensor field of type $(1,1)$. In the context of dimensional reduction one of the components of $e_1^1$ is the KK vector denoted by $A_1^1$. This gauge field has a magnetic dual $A_7^1$ which is a $(1,7)$ mixed symmetry tensor.

 In Table  \ref{dualfieldsandbranes} we present the magnetic duals for degree 2 tensor fields plus the dilaton, and the associated branes.
 The NS5, KKM and $5_2^2$ exhaust the branes that the NSNS sector (save the dilaton) can couple to using 
 relations like (\ref{magneticdual}). Including the dilaton, one more NSNS state is included. This is the S-dual of the D7-brane, denoted here as NS7.
In Ref. \cite{defect2} it was argued that mixed symmetry tensor fields of rank
\beq\label{mixedcouplings}
(n,6+n)\quad \text{for} \quad n=0,1,2,3,4~,
\eeq
are needed to provide the couplings for all solitonic branes, which are 5-branes that have tension proportional to $g_s^{-2}$. They are related by the T-duality chain
\beq\label{fives}
\text{NS5 }(5_2) \quad\longleftrightarrow\quad \text{KKM }(5_2^1) \quad\longleftrightarrow\quad 5_2^2 \quad\longleftrightarrow\quad 5_2^3 \quad\longleftrightarrow\quad 5_2^4~,
\eeq
where the T-duality is always performed along a transverse direction with respect to the dimensionally reduced theory. The 
two branes on the right side of this chain, the $5_2^3$ and the $5_2^4$, are additional solitonic branes that did not appear in Eq. (\ref{branechain}). However, inspection of
Table \ref{dualfieldsandbranes} leads to the observation that potentials coupling to these two branes do not appear. 
Let us explain the reason for their absence. According to (\ref{mixedcouplings}) the $5_2^3$ should couple to a mixed symmetry tensor of type $(3,9)$ which has a field strength of type $(3,10)$. We can schematically write this as $(3,10) = \dd (3,9)$. Dualizing this field we get a pure 3-vector,
\[
R^3 = (3,0) = \star \dd (3,9)~,
\]
which can be associated to the R-flux that appears in non-geometric string compactifications. This is argued as follows.
Obviously $R^3$ cannot be given in terms of a potential in the standard way. In generalized geometry (and also in double field theory \cite{dft1,dft2}), there is a derivation operation that effectively adds one vector index to a given tensor. 
It is enough for our purposes to think of this operation simply as action with the operator 
${\cal D}^i=\beta^{ij}\partial_j$, where $\beta^{ij}$ are the components of the 2-vector $\beta^2$. Then $R^{ijk}={\cal D}^{[i}\beta^{jk]}$, which is the component version of the 
3-vector structure on a quasi-Poisson manifold \cite{quasi}.
Having no associated potential, the $5_2^3$-brane is similar to the type IIA domain wall D8-brane. The D8-brane couples to $C_9$ with field strength $F_{10}$, the Hodge dual of the type IIA Romans mass which is obviously not given as an exterior derivative of a RR potential.
By the same logic the $5^4_2$-brane is similar to the spacetime filling D9-brane. The $5^4_2$ couples to a mixed symmetry tensor of type $(4,
10)$. In ten dimensions such a mixed symmetry tensor does not even have a field strength and is therefore non-dynamical, just like the $C_{10}$ that couples to the D9-brane. We conclude that it is the similarity of these two additional branes in (\ref{fives}) with 
the D8- and D9-branes that accounts for the absence from Table 1 of the tensor fields they couple to. These tensor fields are exotic analogs of $C_9$ and $C_{10}$, which are anyway special potentials in type IIA and IIB respectively.

\begin{table}
\caption{\label{dualfieldsandbranes}Tensor degrees for $\star\dd (\text{Field})\sim \dd(\text{Magnetic Dual)}$ 
and associated branes.}
\label{tab:2}
\begin{tabular}{@{}rrrrl@{}}
\hline
Field & $\dd$ & $\star\dd$ &Magnetic Dual&Associated Brane
\\
\hline
$B_2=(0,2)$ & $(0,3)$&$(0,7)$ &$B_6=(0,6)$&NS5
\\
$\beta^2=(2,0)$ & $(2,1)$&$(2,9)$ &$\beta_8^2=(2,8)$&$5^2_2$
\\
$A_1^1=(1,1)$ &$(1,2)$&$(1,8)$&$A_7^1=(1,7)$&KKM5
\\
$\phi=(0,0)$&$(0,1)$&$(0,9)$&$(0,8)$&NS7
\\
\hline
\end{tabular}
\end{table}

We can make a similar analysis of even heavier branes than the solitonic ones. First we consider the following T-duals of the NS7-brane
\[
\text{NS7 }(7_3) \quad\longleftrightarrow\quad 7_3^1 \quad\longleftrightarrow\quad 7_3^2~,
\]
which couple to mixed symmetry tensors of degree
\beq\label{mixedcouplings2}
(n,8+n)\quad \text{for} \quad n=0,1,2~,
\eeq
where $(0,8)$ is the magnetic dual of the dilaton. As for the $5_2^3$-brane, the $7_3^1$ couples to a mixed symmetry 
tensor with nine form indices. The standard magnetic dual of such an object does not exist but one may give an expression using the derivative operator discussed above, schematically 
$
\star \dd (1,9) = (1,0) = {\cal D}(0,0)~.
$
The $7_3^2$ couples to a mixed symmetry tensor with ten form indices, 
which by the above logic is not a dynamical field. The same is true for the NS9-brane ($9_4$),
which couples to a pure 10-form, S-dual to the RR potential $C_{10}$. Many more branes can be obtained by wrapping the seven- and nine-branes around isometry directions, and performing dualities (see also Ref. \cite{wrap}). 

The presence of the above branes in string theory leads to modified Bianchi identities for the tensors that appear in 
(\ref{extfluxchain}). For $Q_1^2$ the modified Bianchi identity can be determined by taking the type IIB supergravity action together with the $5_2^2$ coupling
\[
S = S_\text{NSNS} + \mu \int \iota_y\iota_z \beta_8^{yz}~,
\]
where $\mu$ is the tension of the $5_2^2$-brane and $S_\text{NSNS}$ is the NSNS action. In order to vary this action, 
one possible approach is to rewrite
the NSNS sector in 
a different set of variables \cite{andriot,Blumenhagen:2013aia}, whence the variation of the action with respect to $\beta^2$ is straightforward and gives
\beq\label{modifiedbianchi}
\dd\left( Q_1^{MN} g_{My} \dd y \w g_{Nz}\dd z\right) = \mu \delta_4~,
\eeq
where $\delta_4$ is a 4-form with support on the world volume of the brane and components transverse to it.  
Eq. \eqref{modifiedbianchi} provides the final step in showing that the $5_2^2$-brane sources the non-geometric $Q_1^2$ flux. 

The main messages of this work can be summarized in the following three points:
\begin{itemize}
\item There exists a plethora of branes aside the standard well-known ones, due to U-duality.
\item These U-dual branes couple to gauge potentials which are mixed symmetry tensors. 
Those tensors are magnetic duals of degree 2 tensor 
fields for branes of co-dimension 2 and higher. On the other hand, for co-dimension 
0 and 1 branes the corresponding mixed symmetry tensors do not possess magnetic duals.
This is in complete analogy to what happens for D8- and D9-branes in type II superstrings.
\item Some exotic branes act as sources of non-geometric fluxes and they lead to modified couplings and Bianchi identities.
\end{itemize}
Further study of the properties of exotic branes is expected to be very useful in gaining a more complete understanding of string vacua both at a conceptual level as well as for phenomenological applications.

\begin{acknowledgement}
Part of this contribution is based on work done in collaboration with George Moutsopoulos and Marco Zagermann.
\end{acknowledgement}

%


\begin{thebibliography}{[1]}
 
 
 \bibitem{defect1}
  E.~A.~Bergshoeff and F.~Riccioni,
  JHEP {\bf 1105} (2011) 131.
  
  \bibitem{defect2}
  E.~A.~Bergshoeff, T.~Ortin and F.~Riccioni,
  Nucl.\ Phys.\ B {\bf 856} (2012) 210.
 
\bibitem{exotic1}
  J.~de Boer and M.~Shigemori,
  Phys.\ Rev.\ Lett.\  {\bf 104} (2010) 251603.
 
 \bibitem{exotic2}
  J.~de Boer and M.~Shigemori,
  Phys.\ Rept.\  {\bf 532} (2013) 65.
  
\bibitem{axel}
  A.~Kleinschmidt,
  JHEP {\bf 1110} (2011) 144.
  
  
\bibitem{Chatzistavrakidis:2013jqa}
  A.~Chatzistavrakidis, F.~F.~Gautason, G.~Moutsopoulos and M.~Zagermann,
  Phys.\ Rev.\ D {\bf 89} (2014) 066004.
  
\bibitem{Hassler}
  F.~Ha{\ss}ler and D.~L\"ust,
  JHEP {\bf 1307} (2013) 048.

\bibitem{betz}
  D.~Andriot and A.~Betz,
  arXiv:1402.5972 [hep-th].
  
\bibitem{Kimura:2014upa}
  T.~Kimura, S.~Sasaki and M.~Yata,
  arXiv:1404.5442 [hep-th].

  
\bibitem{hulltownsend}
  C.~M.~Hull and P.~K.~Townsend,
  Nucl.\ Phys.\ B {\bf 438} (1995) 109.

  \bibitem{Hitchin:2004ut} 
  N.~Hitchin,
  Quart.\ J.\ Math.\ Oxford Ser.\  {\bf 54}, 281 (2003).
  
\bibitem{Gualtieri}
  M.~Gualtieri,
  math/0401221 [math-dg].

\bibitem{mixed1}
  T.~Curtright,
  Phys.\ Lett.\ B {\bf 165} (1985) 304.
  
\bibitem{mixed2}
  C.~M.~Hull,
  JHEP {\bf 0109} (2001) 027.
  
\bibitem{mixed3}
  P.~de Medeiros and C.~Hull,
  Commun.\ Math.\ Phys.\  {\bf 235} (2003) 255.
  

\bibitem{dft1}
  C.~Hull and B.~Zwiebach,
  JHEP {\bf 0909} (2009) 099.

\bibitem{dft2}
  O.~Hohm, C.~Hull and B.~Zwiebach,
  JHEP {\bf 1008} (2010) 008.

  \bibitem{quasi}
A.~Alekseev, Y.~Kosmann-Schwarzbach, E.~Meinrenken,
Canad.\ J.\ Math.\ 54 (2002), 3-29.

\bibitem{wrap}
  E.~A.~Bergshoeff and F.~Riccioni,
  Phys.\ Lett.\ B {\bf 704} (2011) 367.
 
\bibitem{andriot}
  D.~Andriot, M.~Larfors, D.~L\"ust and P.~Patalong,
  JHEP {\bf 1109} (2011) 134.

\bibitem{Blumenhagen:2013aia}
  R.~Blumenhagen, A.~Deser, E.~Plauschinn, F.~Rennecke and C.~Schmid,
  Fortsch.\ Phys.\  {\bf 61} (2013) 893.




 
\end{thebibliography}
\end{document}